\documentclass{article}
\usepackage{amsthm,amsmath,amssymb}
\usepackage{amscd}
\usepackage{graphicx}
\usepackage{mathptmx}
\usepackage{latexsym}
\RequirePackage{fix-cm}
\usepackage{hyperref}

\begin{document}
\title{Work Extracting From Nonextensive  Small System With Feedback and Second Law-Like Inequalities with Quantum Tsallis Entropy}
\author{Saman Amiri\thanks{Contact e-mail: \href{amiri.saman69@gmail.com}{Saman.amiri@shirazu.ac.ir}}~$^{1}$,
Mahdi Mirzaee\thanks{Contact e-mail: \href{m-mirzaee@araku.ac.ir}{m-mirzaee@araku.ac.ir}}~$^{1}$~and
Mohammad Mazhari\thanks{Contact e-mail: \href{m.mazhari@shirazu.ac.ir}{m.mazhari@shirazu.ac.ir}}~$^{2}$\hspace{2.3cm}
\\
\hspace*{-1.95cm}
$^{1}$Department of Physics, Faculty of Sciences, University of Arak, Arak 8349-8-8349, Iran
\\
\hspace*{-2cm}
$^{2}$Department of Physics, College of Sciences, University of Shiraz, Shiraz 71454, Iran}

\maketitle
\begin{abstract}
	
Gibbs-Boltzmann entropy leads to systems that have a strong dependence on initial conditions. In reality most materials behave quite independently of initial conditions. Nonextensive entropy or Tsallis entropy leads to nonextensive statistical mechanics.\, In this paper we calculate Tsallis form of Clausius inequality and then determind upper bound for extracting work from small system in Nonextensive statistical mechanic with mutual information. In the following we
extract mutual information and adjust Maxwell’s demon with quantum feedback
control.

\end{abstract}

\section{Introduction}

Among a large number of studies conducted on the
relationship between thermodynamics and information
processing, particularly provoking is the work of
Szilard who argued that positive work $W_{ext}$ can be
extracted from an isothermal cycle if Maxwell’s demon
plays the role of a feedback controller.\, It is now well understood that the role of the demon does not contradict the second law of thermodynamics, because the initialization of the demon’s memory entails heat dissipation.\, We note that, in the case of an isothermal process, the second law of thermodynamics can be expressed as

\begin{equation}
W_{ext}\leq-\Delta{F^S}
\end{equation}
where $\Delta{F^S}$ is the difference in the Helmholtz free energy between the initial and final thermodynamic equilibrium states.

In a different context, quantum feedback control has attracted considerable attention for controlling and stabilizing a quantum system.\, It can be applied, for example, to squeezing an electromagnetic field, spin squeezing, and stabilizing macroscopic coherence.\, While the theoretical framework of quantum feedback control as a stochastic dynamic system is well developed, the possible thermodynamic gain of quantum
feedback control has yet to be fully understood.\\

In this Letter, we derive a new thermodynamic inequality which sets the fundamental limit on the work that can be extracted from multi-heat-baths with discrete quantum feedback control, consisting of quantum measurement and a mechanical operation depending on the measurement outcome.\, The maximum work is characterized by a generalized mutual information content between the thermodynamic system and the feedback controller.
We consider a thermodynamic process for system S
which can contact heat baths $B_{1}$; $B_{2}$; . . . ; $B_{n}$ at respective temperatures $T_{1}$; $T_{2}$; . . . ; $T_{n}$.\, We assume that system S is in
thermodynamic equilibrium in the initial and final states. For simplicity, we also assume that the initial and final temperature of S is given by $T={{\left( {{k}_{B}}\beta  \right)}^{-1}}$. This can be
realized by contacting S with, for example, B1 in the
preparation of the initial state and during equilibration to the final state; in this case $T={{T}_{1}}$.\, We do not, however, assume that the system is in thermodynamic equilibrium between the initial and final states. Then the second law of thermodynamics is expressed as the Clausius inequlity:
\begin{equation}
\sum\limits_{m}{{{\beta }_{m}}{{Q}_{m}}\le 0}
\end{equation}

\section{Nonextensive Concepts}

Methods of nonextensive statistical mechanics have found use in
much many topics of physics and other sciences.\, In the nonextensive statistical physics, the Tsallis q-entropy of the probability distribution p is defined as

\begin{equation}
{{H}_{q}}(p)=\frac{1}{q-1}\left( \sum\limits_{i=1}^{m}{p_{i}^{q}-1} \right)=-\sum\limits_{i=1}^{m}{p_{i}^{q}{{\ln }_{q}}{{p}_{i}}}
\end{equation}
where ${{\ln }_{q}}x=\frac{{{x}^{1-q}}-1}{1-q}$ is q-logarithm function.The quantum Tsallis entropy of density operator $\rho$ is defined as:
\begin{equation}
{{S}_{q}}(\hat\rho )=\frac{1}{1-q}tr\left( {{\hat\rho }^{q}}-\hat\rho  \right)
\end{equation}
For $q = 1$, we have the von Neumann entropy $S(\hat\rho )=-tr(\hat\rho \ln \hat\rho )$.
 
\section{Q-functions in Canonical Ensemble}

Canonical equilibrium distribution in nonextensive statistical is defined as:
\begin{equation}
{{\hat\rho }_{q}}=\frac{e_{q}^{-{\beta }'\hat{H}}}{{{{\bar{Z}}}_{q}}}
\end{equation}
where in equation (5)
\begin{equation}
{{\bar{Z}}_{q}}=tr\left( e_{q}^{-{\beta }'\hat{H}} \right)
\end{equation}
and
\begin{equation}
{\beta }'=\frac{\beta }{\sum\nolimits_{i=1}^{w}{{{\left( {{p}_{i}} \right)}^{q}}+\left( 1-q \right)\beta {{U}_{q}}}}
\end{equation}
With the canonical distribution, free energy $F^S$ can be calculated as
\begin{equation}
{{F}_{q}}=-{{k}_{B}}T{{\ln }_{q}}{{Z}_{q}}
\end{equation}
where
\begin{equation}
{{\ln }_{q}}{{Z}_{q}}={{\ln }_{q}}{{\bar{Z}}_{q}}-\beta {{U}_{q}}
\end{equation}
The average of measurement outcomes of observable A is given by
\begin{equation}
{{\left\langle A \right\rangle }_{q}}=tr\left( {{\hat\rho }^{q}}A \right)
\end{equation}

\section{Q-Clausius inequality}
We consider a thermodynamic process of system S that can contact heat baths  $B_{1}$; $B_{2}$; . . . ; $B_{n}$ and we assume that the total of S and $B_{m}$’s obeys a unitary evolution.\, The total Hamiltonian can be written as
\begin{equation}
\hat{H}(\lambda ,\left\{ {{c}_{m}} \right\})={{\hat{H}}^{S}}(\lambda )+\sum\limits_{m=1}^{n}{({{{\hat{H}}}^{S{{B}_{m}}}}({{c}_{m}})+{{{\hat{H}}}^{{{B}_{m}}}})}
\end{equation}
where ${\hat{H}}^{S}(\lambda )$ is the Hamiltonian of S, ${{{\hat{H}}}}^{S{{B}_{m}}}({{c}_{m}})$ is the interaction Hamiltonian between S and ${B}_{m}$, and ${{{\hat{H}}^{{B}_{m}}}}$ is the Hamiltonian of ${B}_{m}$. Here, $\lambda$ describes controllable
external parameters, and ${c}_{m}$ describes external parameters to control the interaction
between S and ${B}_{m}$.

We assume that the initial state of the total system is given by
\begin{equation}
{{\hat\rho }_{i}}=\hat\rho _{i}^{s}\otimes \hat\rho _{q}^{{{B}_{1}}}\otimes ...\otimes \hat\rho _{q}^{{{B}_{n}}}
\\
\end{equation}
The process with unitary evolution of the total system, $\hat{U}\equiv T\exp \left( -i\int\limits_{0}^{\tau }{\hat{H}(\lambda (t),\left\{ {{c}_{m}}(t) \right\})dt} \right)$ leads to the final state
\begin{equation}
{{\hat\rho }_{f}}\equiv \hat{U}{{\hat\rho }_{i}}{\hat{U}^{\dagger }}
\end{equation}
Then the Tsallis entropy of initial state of total system is given by
\begin{equation}
\begin{aligned}
& {{S}_{q}}({{\hat\rho }_{i}})=-tr\left( \hat\rho _{i}^{q}{{\ln }_{q}}{{\hat\rho }_{i}} \right)={{S}_{q}}(\hat\rho _{i}^{s})tr(\hat\rho _{q}^{q{{B}_{1}}})...tr(\hat\rho _{q}^{q{{B}_{n}}})+ \\
& \sum\limits_{m}{\frac{{{{{\beta }'}}_{m}}}{{{{\bar{Z}}}_{m}}^{1-q}}\left( tr\left[ \hat\rho _{q}^{q{{B}_{m}}}{\hat{H}^{{{B}_{m}}}} \right]+\frac{1}{{{{{\beta }'}}_{m}}}tr\left( \hat\rho _{q}^{q{{B}_{m}}}{{\ln }_{q}}{{Z}_{mq}} \right)+\frac{{{\beta }_{m}}}{{{{{\beta }'}}_{m}}}tr\left( \hat\rho _{q}^{q{{B}_{m}}}{{U}_{mq}} \right) \right)} \\ 
& tr(\hat\rho _{i}^{sq})-(1-q)\left[ tr\left( \hat\rho _{i}^{sq}{{\ln }_{q}}{{\hat\rho }_{i}}\otimes \hat\rho _{q}^{q{{B}_{1}}}{{\ln }_{q}}\hat\rho _{q}^{{{B}_{1}}}\otimes ...\otimes \hat\rho _{q}^{q{{B}_{m}}}{{\ln }_{q}}\hat\rho _{q}^{{{B}_{m}}} \right) \right] \\ 
\end{aligned}
\end{equation}
From Klein’s inequality, we also have
\begin{equation}
\begin{aligned}
& {{S}_{q}}({{\hat\rho }_{f}})\le -tr\left( \hat\rho _{f}^{q}{{\ln }_{q}}(\hat\rho _{f}^{s}\otimes \hat\rho _{q}^{{{B}_{1}}}\otimes ...\otimes \hat\rho _{q}^{{{B}_{n}}} \right)= \\ 
& -tr\left( \hat\rho _{f}^{q}{{\ln }_{q}}{{\hat\rho }_{f}} \right)={{S}_{q}}(\hat\rho _{f}^{s})tr(\hat\rho _{q}^{q{{B}_{1}}})...tr(\hat\rho _{q}^{q{{B}_{n}}})+ \\ 
& \sum\limits_{m}{\frac{{{{{\beta }'}}_{m}}}{{{{\bar{Z}}}_{m}}^{1-q}}\left( tr\left[ \hat\rho _{q}^{q{{B}_{m}}}{\hat{H}^{{{B}_{m}}}} \right]+\frac{1}{{{\beta }'}}tr\left( \hat\rho _{q}^{q{{B}_{m}}}{{\ln }_{q}}{{Z}_{mq}} \right)+\frac{{{\beta }_{m}}}{{{{{\beta }'}}_{m}}}tr\left( \hat\rho _{q}^{q{{B}_{m}}}{{U}_{mq}} \right) \right)} \\ 
& tr(\hat\rho _{f}^{sq})-(1-q)\left[ tr\left( \hat\rho _{i}^{sq}{{\ln }_{q}}{{\rho }_{i}}\otimes \hat\rho _{q}^{q{{B}_{1}}}{{\ln }_{q}}\hat\rho _{q}^{{{B}_{1}}}\otimes ...\otimes \hat\rho _{q}^{q{{B}_{m}}}{{\ln }_{q}}\hat\rho _{q}^{{{B}_{m}}} \right) \right] \\ 
\end{aligned}
\end{equation}
Therefore we obtain
\begin{equation}
\begin{aligned}
& \left( {{S}_{q}}(\hat\rho _{f}^{s})-{{S}_{q}}(\hat\rho _{i}^{s}) \right)tr(\hat\rho _{q}^{q{{B}_{1}}})...tr(\hat\rho _{q}^{q{{B}_{n}}})\ge  \\ 
& \sum\limits_{m}{\frac{{{{{\beta }'}}_{m}}}{{{{\bar{Z}}}_{m}}^{1-q}}\left[ tr\left[ \hat\rho _{i}^{q}{\hat{H}^{{{B}_{m}}}} \right]-tr\left[ \hat\rho _{f}^{q}{\hat{H}^{{{B}_{m}}}} \right] \right]}+ \\ 
& \sum\limits_{m}{\frac{1}{{{{{\beta }'}}_{m}}}\left[ tr\left( \hat\rho _{i}^{q}{{\ln }_{q}}{{Z}_{mq}} \right)-tr\left( \hat\rho _{f}^{q}{{\ln }_{q}}{{Z}_{mq}} \right) \right]}+ \\ 
& \sum\limits_{m}{\frac{{{\beta }_{m}}}{{{{{\beta }'}}_{m}}}\left[ tr\left( \hat\rho _{i}^{q}{{U}_{mq}} \right)-tr\left( \hat\rho _{f}^{q}{{U}_{mq}} \right) \right]}+ \\ 
& (1-q)\left[ {{S}_{q}}({{\hat\rho }_{{{B}_{1}}}})...{{S}_{q}}({{\hat\rho }_{{{B}_{m}}}}) \right]\left[ {{S}_{q}}(\hat\rho _{f}^{s})-{{S}_{q}}(\hat\rho _{i}^{s}) \right] \\ 
\end{aligned}
\end{equation}
and in summary
\begin{equation}
\begin{aligned}
& \left( {{S}_{q}}(\hat\rho _{f}^{s})-{{S}_{q}}(\hat\rho _{i}^{s}) \right)tr(\hat\rho _{q}^{q{{B}_{1}}})...tr(\rho _{q}^{q{{B}_{n}}})\ge  \\ 
& \sum\limits_{m}{\frac{{{{{\beta }'}}_{m}}}{{{{\bar{Z}}}_{m}}^{1-q}}{{Q}_{qm}}}+\sum\limits_{m}{\frac{{{\beta }_{m}}}{{{{{\beta }'}}_{m}}}\left[ tr(\hat\rho _{i}^{q})-tr(\hat\rho _{f}^{q}) \right]\left( F_{mq}^{B}-U_{mq}^{B} \right)} \\ 
& +(1-q)\left[ {{S}_{q}}({{\hat\rho }_{{{B}_{1}}}})...{{S}_{q}}({{\hat\rho }_{{{B}_{m}}}}) \right]\left( {{S}_{q}}(\hat\rho _{f}^{s})-{{S}_{q}}(\hat\rho _{i}^{s}) \right) \\ 
\end{aligned}
\end{equation}
where
\begin{equation}
{{Q}_{qm}}=\left[ tr\left( \hat\rho _{i}^{q}{\hat{H}^{mB}} \right)-tr\left( \rho _{f}^{q}{\hat{H}^{mB}} \right) \right].
\end{equation}
is the heat that is absorbed by system S from heat bath ${B}_{m}$ and ${{U}_{mq}}=-\frac{\partial }{\partial \beta }{{\ln }_{q}}{{Z}_{mq}}$ is internal energy of baths and third term in equation (17) determind interaction between system and baths.
Note for $q=1$ Inequality q-Clausius becomes standard inequality Clausius.

Isothermal process. We next consider the case in which there is a single heat bath.\, We then assume that the initial state of S is the canonical distribution as
\begin{equation}
\hat\rho _{iq}^{s}=\frac{e_{q}^{-{\beta }'\hat{H}_{i}^{s}}}{{{{\bar{Z}}}_{q}}}
\end{equation}
where $\bar{Z}_{iq}^{s}=tr\left( e_{q}^{-{\beta }'\hat{H}_{i}^{s}} \right)$.We also introduce notations as
\begin{equation}
F_{iq}^{s}=-{{k}_{B}}T{{\ln }_{q}}Z_{iq}^{s}
\end{equation}
\begin{equation}
\hat\rho _{fq}^{s}=\frac{e_{q}^{-{\beta }'\hat{H}_{f}^{s}}}{{{{\bar{Z}}}_{q}}}, \bar{Z}_{fq}^{s}=tr\left( e_{q}^{-{\beta }'\hat{H}_{f}^{s}} \right),  F_{fq}^{s}=-{{k}_{B}}T{{\ln }_{q}}Z_{fq}^{s}
\end{equation}
Then the Tsallis entropy of initial and final state of  system is given by
\begin{equation}
\begin{aligned}
& {{S}_{q}}\left( \hat\rho _{iq}^{s} \right)=-tr\left( \hat\rho _{iq}^{sq}{{\ln }_{q}}{{\hat\rho }_{iq}} \right)= \\ 
& \frac{{{\beta }'}}{\bar{Z}_{iq}^{1-q}}\left[ tr\left( \hat\rho _{iq}^{sq}\hat{H}_{i}^{s} \right)+\frac{1}{{{\beta }'}}tr\left( \rho _{iq}^{sq}{{\ln }_{q}}Z_{iq}^{s} \right)+\frac{\beta }{{{\beta }'}}tr\left( \hat\rho _{iq}^{sq}U_{iq}^{s} \right) \right] \\ 
\end{aligned}
\end{equation}
and
\begin{equation}
{{S}_{q}}\left( \hat\rho _{fq}^{s} \right)=\frac{{{\beta }'}}{\bar{Z}_{fq}^{1-q}}\left[ tr\left( \hat\rho _{fq}^{sq}\hat{H}_{f}^{s} \right)+\frac{1}{{{\beta }'}}tr\left( \hat\rho _{fq}^{sq}{{\ln }_{q}}Z_{fq}^{s} \right)+\frac{\beta }{{{\beta }'}}tr\left( \hat\rho _{fq}^{sq}U_{fq}^{s} \right) \right]
\end{equation}
From Klein’s inequality, we obtain
\begin{equation}
\begin{aligned}
& {{S}_{q}}\left( \hat\rho _{fq}^{s} \right)-{{S}_{q}}\left( \hat\rho _{iq}^{s} \right)\le \frac{{{\beta }'}}{\bar{Z}_{fq}^{1-q}}E_{f}^{s}-\frac{{{\beta }'}}{\bar{Z}_{iq}^{1-q}}E_{i}^{s}-\frac{\beta }{\bar{Z}_{fq}^{1-q}}F_{fq}^{s}tr\left( \hat\rho _{fq}^{sq} \right)+ \\ 
& \frac{\beta }{\bar{Z}_{iq}^{1-q}}F_{iq}^{s}tr\left( \hat\rho _{iq}^{sq} \right)+\frac{\beta }{\bar{Z}_{fq}^{1-q}}U_{fq}^{s}tr\left( \hat\rho _{fq}^{sq} \right)-\frac{\beta }{\bar{Z}_{iq}^{1-q}}U_{iq}^{s}tr\left( \hat\rho _{iq}^{sq} \right) \\ 
\end{aligned}
\end{equation}
where 
\begin{equation}
E_{i}^{s}=tr\left( \hat\rho _{iq}^{sq}\hat{H}_{i}^{s} \right)
\end{equation}
and 
\begin{equation}
E_{f}^{s}=tr\left( \hat\rho _{fq}^{sq}\hat{H}_{f}^{s} \right)
\end{equation}
are energy stored in the system.\,that is clear for $q=1$ equation (24) becomes the standard form as follow
\begin{equation}
S\left( \hat\rho _{f}^{s} \right)-S\left( \hat\rho _{i}^{s} \right)\le \beta \left( \Delta {{E}^{s}}-\Delta {{F}^{s}} \right)
\end{equation} 

\section{Entropy-Mutual Information Inequality}
We derive the entropy balance of a quantum system that obeys a quantum measurement and quantum feedback control in addition to unitary evolutions.
Let ${{\hat{\rho }}_{i}}$ be an arbitrary initial density operator of quantum system,\,which evolves as follows.\\

Step 1: Unitary evolution. From time 0 to $t_1$, the system undergoes unitary evolution ${{\hat{U}}_{i}}$. At time t1, the density operator is given by ${{\hat{\rho }}_{1}}={{\hat{U}}_{i}}{{\hat{\rho }}_{i}}\hat{U}_{i}^{\dagger }$.

Step 2: Measurement. From time $t_1$ to $t_2$, we perform a quantum measurement on the system. We assume that the measurement is described by measurement operators
$\left\{ {{{\hat{M}}}_{k}} \right\}$ with k’s being measurement outcomes, which leads to POVM
\begin{equation}
{{\hat{E}}_{k}}\equiv \hat{M}_{k}^{\dagger }{{\hat{M}}_{k}}
\end{equation}
The post-measurement state corresponding to outcome k is given by
\begin{equation}
\hat{\rho }_{2}^{(k)}\equiv \frac{1}{{{p}_{k}}}{{\hat{M}}_{k}}{{\hat{\rho }}_{1}}\hat{M}_{k}^{\dagger }
\end{equation}
and the ensemble average is given by
\begin{equation}
{{\hat{\rho }}_{2}}=\sum\limits_{k}{{{p}_{k}}\hat{\rho }_{2}^{\left( k \right)}}=\sum\limits_{k}{{{{\hat{M}}}_{k}}{{{\hat{\rho }}}_{1}}\hat{M}_{k}^{\dagger }}
\end{equation}

Step 3: Feedback control. From  $t_2$ to  $t_3$, we perform feedback control; the corresponding unitary operator ${{\hat{U}}_{k}}$ depends on measurement outcome k. The postfeedback state corresponding to outcome k is given by
\begin{equation}
\hat{\rho }_{3}^{(k)}={{\hat{U}}_{k}}\hat{\rho }_{2}^{(k)}\hat{U}_{k}^{\dagger }=\frac{1}{{{p}_{k}}}{{\hat{U}}_{k}}{{\hat{M}}_{k}}{{\hat{\rho }}_{1}}\hat{M}_{k}^{\dagger }\hat{U}_{k}^{\dagger }
\end{equation}
and the ensemble average is given by
\begin{equation}
{{\hat{\rho }}_{3}}=\sum\limits_{k}{{{p}_{k}}{{{\hat{U}}}_{k}}\hat{\rho }_{2}^{(k)}\hat{U}_{k}^{\dagger }}=\sum\limits_{k}{{{{\hat{U}}}_{k}}{{{\hat{M}}}_{k}}{{{\hat{\rho }}}_{1}}\hat{M}_{k}^{\dagger }\hat{U}_{k}^{\dagger }}
\end{equation}

Step 4: Unitary evolution. After the feedback, from time $t_3$ to $\tau $, the system evolves
according to unitary operator ${{\hat{U}}_{f}}$ which is independent of outcome k. The final state is ${{\hat{\rho }}_{f}}={{\hat{U}}_{f}}{{\hat{\rho }}_{3}}{{\hat{U}}_{f}}$.

The difference in the Tsallis entropy S between the initial and final states can be bounded as follows:
\begin{equation}
\begin{aligned}
& {{S}_{q}}({{{\hat{\rho }}}_{i}})-{{S}_{q}}({{{\hat{\rho }}}_{f}}) \\ 
& ={{S}_{q}}({{{\hat{\rho }}}_{1}})-{{S}_{q}}({{{\hat{\rho }}}_{3}}) \\ 
& \le {{S}_{q}}({{{\hat{\rho }}}_{1}})-\sum\limits_{k}{{{p}_{k}}}{{S}_{q}}(\hat{\rho }_{3}^{(k)}) \\ 
& ={{S}_{q}}({{{\hat{\rho }}}_{1}})-\sum\limits_{k}{{{p}_{k}}}{{S}_{q}}(\hat{\rho }_{2}^{(k)}) \\ 
& ={{S}_{q}}({{{\hat{\rho }}}_{1}})+\sum\limits_{k}{p_{k}^{1-q}tr\left( \hat{E}_{k}^{\frac{q}{2}}{{{\hat{\rho }}}_{1}}\hat{E}_{k}^{\frac{q}{2}}{{\ln }_{q}}\frac{\sqrt{{\hat{E}}}{{{\hat{\rho }}}_{1}}\sqrt{{\hat{E}}}}{{{p}_{k}}} \right)} \\ 
& ={{S}_{q}}({{{\hat{\rho }}}_{1}})+{{H}_{q}}(p)+\sum\limits_{k}{tr}\left( \hat{E}_{k}^{\frac{q}{2}}{{{\hat{\rho }}}_{1}}\hat{E}_{k}^{\frac{q}{2}}{{\ln }_{q}}\sqrt{{\hat{E}}}{{{\hat{\rho }}}_{1}}\sqrt{{\hat{E}}} \right) \\ 
\end{aligned}\
\end{equation}
 From the definition of
QC-mutual information [22], we obtain
\begin{equation}
{{S}_{q}}({{\hat{\rho }}_{i}})-{{S}_{q}}({{\hat{\rho }}_{f}})\le I_{q}^{QC}\
\end{equation}
Result by equation (17)(24)(34):
\begin{equation}
\begin{aligned}
& {\left( \sum\limits_{m}{\frac{{{{{\beta }'}}_{m}}}{{{{\bar{Z}}}_{m}}^{1-q}}{{Q}_{qm}}} \right)}/{tr(\hat\rho _{q}^{q{{B}_{1}}})...tr(\hat\rho _{q}^{q{{B}_{n}}})}\;-{\beta }'\left\{ \frac{E_{f}^{s}}{\bar{Z}_{fq}^{1-q}}-\frac{E_{i}^{s}}{\bar{Z}_{iq}^{1-q}} \right\}\le  \\ 
& +\beta \left\{ -\frac{1}{\bar{Z}_{fq}^{1-q}}F_{fq}^{s}tr\left( \hat\rho _{fq}^{sq} \right)+\frac{1}{\bar{Z}_{iq}^{1-q}}F_{iq}^{s}tr\left( \rho _{iq}^{sq} \right) \right\} \\ 
& +\beta \left\{ \frac{1}{\bar{Z}_{fq}^{1-q}}U_{fq}^{s}tr\left( \hat\rho _{fq}^{sq} \right)-\frac{1}{\bar{Z}_{iq}^{1-q}}U_{iq}^{s}tr\left( \hat\rho _{iq}^{sq} \right) \right\} \\ 
& {-\sum\limits_{m}{\frac{{{\beta }_{m}}}{{{{{\beta }'}}_{m}}}\left[ tr(\hat\rho _{i}^{q})-tr(\hat\rho _{f}^{q}) \right]\left( F_{mq}^{B}-U_{mq}^{B} \right)}}/{tr(\hat\rho _{q}^{q{{B}_{1}}})...tr(\hat\rho _{q}^{q{{B}_{n}}})}\; \\ 
& +{(1-q)\left[ {{S}_{q}}({{\rho }_{{{B}_{1}}}})...{{S}_{q}}({{\hat\rho }_{{{B}_{m}}}}) \right]\left( {{S}_{q}}(\hat\rho _{f}^{s})-{{S}_{q}}(\hat\rho _{i}^{s}) \right)}/{tr(\hat\rho _{q}^{q{{B}_{1}}})...tr(\hat\rho _{q}^{q{{B}_{n}}})}\;+I_{q}^{QC} \\ 
\end{aligned}\
\end{equation}
from first law thermodynamics, work extraction from system by feedback control in nonextensive statistical is given by
\begin{equation}
\begin{aligned}
& W_{ext}^{S}\le -\beta \Delta F_{q}^{s}+\beta \Delta U_{q}^{s}{-\sum\limits_{m}{\frac{{{\beta }_{m}}}{{{{{\beta }'}}_{m}}}\left[ tr(\hat\rho _{i}^{q})-tr(\hat\rho _{f}^{q}) \right]\left( F_{mq}^{B}-U_{mq}^{B} \right)}}/{tr(\hat\rho _{q}^{q{{B}_{1}}})...tr\hat(\hat\rho _{q}^{q{{B}_{n}}})}\; \\ 
& +{(1-q)\left[ {{S}_{q}}({{\hat\rho }_{{{B}_{1}}}})...{{S}_{q}}({{\hat\rho }_{{{B}_{m}}}}) \right]\left( {{S}_{q}}(\hat\rho _{f}^{s})-{{S}_{q}}(\hat\rho _{i}^{s}) \right)}/{tr(\hat\rho _{q}^{q{{B}_{1}}})...tr(\hat\rho _{q}^{q{{B}_{n}}})}\;+I_{q}^{QC} \\ 
\end{aligned}
\end{equation}

\section{Erasure Process}
We consider the following process for the information erasure in the presence of a
single heat bath. The pre-erasure state means the postmeasurement state, in which the memory stores the information of the measured
system. In the pre-erasure state, M stores outcome “k” with probability $p_k$.

We assume that, before the information erasure, the state of M under the condition
of "k" is in the canonical distribution ${{\hat{\rho }}_{q,k}}$, and that the total pre-erasure state of M
is given by
\begin{equation}
\hat{\rho }_{i}^{M}\equiv \sum\limits_{k}{{{p}_{k}}}\hat{\rho }_{q,k}^{M}\
\end{equation}
We also assume that the initial states of M and B are not correlated, and that the initial state of the total system is given by
\begin{equation}
\hat{\rho }_{i}^{MB}\equiv \hat{\rho }_{i}^{M}\otimes \hat{\rho }_{q,k}^{B}
\end{equation}
We consider the erasure process from $t = 0$ to $t=\tau $. During the erasure process, we change the Hamiltonian of M with a protocol which need to be independent of k.
The total Hamiltonian at time t then is given by
\begin{equation}
{{\hat{H}}^{MB}}(t)={{\hat{H}}^{M}}(t)+{{\hat{H}}^{\operatorname{int}}}(t)+{{\hat{H}}^{B}}
\end{equation}
where ${{\hat{H}}^{\operatorname{int}}}(t)$ is the interaction Hamiltonian between M and B.
The time evolution of the total system from time 0 to  $\tau $ is then given by the unitary operator $\hat{U}\equiv T\exp \left( -i\int{{{{\hat{H}}}^{MB}}(t)dt} \right)$, which gives the post-erasure state of the total systems
\begin{equation}
\hat{\rho }_{f}^{MB}=\hat{U}\hat{\rho }_{i}^{MB}{{\hat{U}}^{\dagger }}\
\end{equation}
We now derive the minimal energy cost that is needed for the erasure process.
From the general second law (17), we obtain
\begin{equation}
\begin{aligned}
& \left( {{S}_{q}}(\rho _{f}^{M})-{{S}_{q}}(\rho _{i}^{M}) \right)tr(\rho _{q}^{qB})\ge  \\ 
& \frac{{{\beta }'}}{\bar{Z}_{B}^{1-q}}Q_{eras}^{M}+\frac{\beta }{{{\beta }'}}\left[ tr(\rho _{i}^{qM})-tr(\rho _{f}^{qM}) \right]\left( F_{q}^{B}-U_{q}^{B} \right) \\ 
& +(1-q)\left( {{S}_{q}}(\rho _{q}^{B}) \right)\left( {{S}_{q}}(\rho _{f}^{M})-{{S}_{q}}(\rho _{i}^{M}) \right) \\ 
\end{aligned}
\end{equation}
where $Q_{eras}^{M}=\left[ tr\left( \rho _{i}^{qMB}{\hat{H}^{B}} \right)-tr\left( \rho _{f}^{qMB}{\hat{H}^{B}} \right) \right]$ is the heat that is absorbed in M during the erasure process. On the other hand, ${{S}_{q}}\left( \hat{\rho }_{i}^{M} \right)$ can be decomposed as
\begin{equation}
\begin{aligned}
& {{S}_{q}}\left( \hat{\rho }_{i}^{M} \right)\equiv tr(\hat{\rho }_{q,k}^{Mq}){\hat{H}_{q}}(p)+\sum\limits_{k}{p_{k}^{q}{{S}_{q}}(\hat{\rho }_{q,k}^{M})} \\ 
& +\sum\limits_{k}{p_{k}^{q}(1-q)({{\ln }_{q}}p_{k}){{S}_{q}}(\hat{\rho }_{q,k}^{M})} \\ 
\end{aligned}
\end{equation}
if the conditional canonical distribution under the
condition of outcome “k” is given by 
\begin{equation}
\rho _{q,k}^{M}=\frac{e_{q}^{-{\beta }'\hat{H}_{k}^{M}}}{{{{\bar{Z}}}_{q,k}}}
\end{equation}
we note that
\begin{equation}
\begin{aligned}
& {{S}_{q}}\left( \hat{\rho }_{q,}^{M} \right)\equiv \frac{{{\beta }'}}{\bar{Z}_{q}^{1-q}}E_{k}^{M}-\frac{\beta }{\bar{Z}_{q}^{1-q}}F_{k}^{M}tr(\hat{\rho }_{q,k}^{Mq}) \\ 
& +\frac{\beta }{\bar{Z}_{q}^{1-q}}U_{k}^{M}tr(\hat{\rho }_{q,k}^{Mq}) \\ 
\end{aligned}
\end{equation}
holds. On the other hand, from Klein’s inequality, we have
\begin{equation}
\begin{aligned}
& {{S}_{q}}\left( \hat{\rho }_{f}^{M} \right)\le -tr(\hat{\rho }_{f}^{Mq}\ln \hat{\rho }_{q,0}^{M})=\frac{{{\beta }'}}{\bar{Z}_{q}^{1-q}}E_{0}^{M}-\frac{\beta }{\bar{Z}_{q}^{1-q}}F_{0}^{M}tr(\hat{\rho }_{f}^{M}) \\ 
& +\frac{\beta }{\bar{Z}_{q}^{1-q}}U_{0}^{M}tr(\hat{\rho }_{f}^{M}) \\ 
\end{aligned}
\end{equation}
where
\begin{equation}
E_{k}^{M}=tr(\hat{\rho }_{q,k}^{Mq}\hat{H}_{k}^{M})\
\end{equation}
and
\begin{equation}
E_{0}^{M}=tr(\hat{\rho }_{f}^{Mq}\hat{H}_{0}^{M})\
\end{equation}
are the internal energies of M. now from equations (41)(42)(44)(45) we obtain
\begin{equation}
\begin{aligned}
& \frac{{{\beta }'}}{\bar{Z}_{q}^{1-q}}\Delta E_{eras}^{M}-\frac{\beta }{\bar{Z}_{q}^{1-q}}\Delta F_{eras}^{M}+\frac{\beta }{\bar{Z}_{q}^{1-q}}\Delta U_{q}^{M} \\ 
& -tr(\hat{\rho }_{q,k}^{Mq}){{H}_{q}}(p)-\sum\limits_{k}{p_{k}^{q}(1-q)({{\ln }_{q}}p_{k}){{S}_{q}}(\hat{\rho }_{q,k}^{M})}\ge  \\ 
& \frac{1}{tr(\hat{\rho }_{q}^{qB})}\left\{ \frac{{{\beta }'}}{\bar{Z}_{B}^{1-q}}Q_{eras}^{M}+\frac{\beta }{{{\beta }'}}\left[ tr(\rho _{i}^{qM})-tr(\rho _{f}^{qM}) \right]\left( F_{q}^{B}-U_{q}^{B} \right) \right\} \\ 
& +\frac{1}{tr(\hat{\rho }_{q}^{qB})}\left\{ (1-q)\left( {{S}_{q}}(\rho _{q}^{B}) \right)\left( {{S}_{q}}(\rho _{f}^{M})-{{S}_{q}}(\rho _{i}^{M}) \right) \right\} \\ 
\end{aligned}
\end{equation}
where
\begin{equation}
\Delta E_{eras}^{M}\equiv tr(\hat{\rho }_{f}^{Mq}H_{0}^{M})-\sum\limits_{k}{{{p}_{k}}tr(\hat{\rho }_{q,k}^{Mq}H_{k}^{M})}
\end{equation}
is the difference of the averaged internal energies of M, and
\begin{equation}
\Delta F_{eras}^{M}\equiv F_{0}^{M}tr(\hat{\rho }_{f}^{qM})-\sum\limits_{k}{p_{k}^{q}F_{k}^{M}tr(\hat{\rho }_{q,k}^{Mq})}
\end{equation}
is the difference of the averaged free energies of M. therefore From the first law of thermodynamics we obtain
\begin{equation}
\begin{aligned}
& W_{eras}^{M}\ge tr(\hat{\rho }_{q,k}^{Mq}){{H}_{q}}(p)+\frac{\beta }{\bar{Z}_{q}^{M\left( 1-q \right)}}\Delta F_{eras}^{M}-\frac{\beta }{\bar{Z}_{q}^{M\left( 1-q \right)}}\Delta U_{q}^{M} \\ 
& +\sum\limits_{k}{p_{k}^{q}(1-q)({{\ln }_{q}}{{p}_{k}}){{S}_{q}}(\hat{\rho }_{q,can,k}^{M})} \\ 
& +\frac{1}{tr(\hat{\rho }_{q}^{qB})}\left\{ \frac{\beta }{{{\beta }'}}\left[ tr(\rho _{i}^{qM})-tr(\rho _{f}^{qM}) \right]\left( F_{q}^{B}-U_{q}^{B} \right) \right\} \\ 
& +\frac{1}{tr(\hat{\rho }_{q}^{qB})}\left\{ (1-q)\left( {{S}_{q}}(\rho _{q}^{B}) \right)\left( {{S}_{q}}(\rho _{f}^{M})-{{S}_{q}}(\rho _{i}^{M}) \right) \right\} \\ 
\end{aligned}
\end{equation}
\section{Measurement Process}
We next consider the measurement processes. Suppose that memory M performs
a measurement on a measured system S, and stores outcome "k" with probability
$p_k$. We assume that the memory is in contact with heat bath $B_M$. On the other hand, during the measurement, measured system S adiabatically evolves or is in contact with a different heat bath, denoted as $B_S$, which
is different from $B_M$. The latter assumption corresponds to the condition that the thermal noises on M and S are independent. The total Hamiltonian is then given by
\begin{equation}
{{\hat{H}}^{tot}}(t)={{\hat{H}}^{M}}(t)+{{\hat{H}}^{M{{B}_{M}}}}(t)+{{\hat{H}}^{{{B}_{M}}}}+{{\hat{H}}^{S}}(t)+{{\hat{H}}^{S{{B}_{S}}}}(t)+{{\hat{H}}^{{{B}_{S}}}}(t)+{{\hat{H}}^{MS}}(t)
\end{equation}
Step 1: Initial state. The initial state of M is in the standard state "0"; we assume
that the initial state of M is the conditional canonical distribution under the condition that the support of the density operator is in $H_{0}^{M}$ .Then the initial density operator of the total system is given by
\begin{equation}
\hat{\rho }_{i}^{tot}=\hat{\rho }_{q,0}^{M}\otimes \hat{\rho }_{q}^{{{B}_{M}}}\otimes {{\hat{\rho }}^{S{{B}_{s}}}}
\end{equation}
Step 2: Unitary evolution. The total system evolves unitarily due to the Hamiltonian. We write
\begin{equation}
{{\hat{U}}_{i}}\equiv T\exp \left( -i\int\limits_{0}^{{{t}_{1}}}{{{{\hat{H}}}^{tot}}(t)dt} \right)
\end{equation}
By this interaction, memory M becomes entangled with measured system S. After
this interaction, the total density operator is given by ${{\hat{U}}_{i}}\hat{\rho }_{i}^{tot}\hat{U}_{i}^{\dagger }$.
We assume that the post-measurement state is given by
\begin{equation}
\hat{\rho }_{f}^{tot}=\sum\limits_{k}{{{{\hat{M}}}_{k}}\hat{\rho }_{i}^{S{{B}_{S}}}\hat{M}_{k}^{\dagger }\otimes }\hat{\rho }_{k}^{M{{B}_{M}}}
\end{equation}
where ${{\hat{M}}_{k}}$ are the measurement operators, and $\hat{\rho }_{k}^{M{{B}_{M}}}$ are the density operators
of M and BM that are mutually orthogonal. Assumption (55) is equivalent to the assumption that any element of the POVM is given by a single measurement operator:
\begin{equation}
{{\hat{E}}_{k}}\equiv \hat{M}_{k}^{\dagger }{{\hat{M}}_{k}}
\end{equation}
Tsallis entropy is invariant under unitary evolutions and increases under projections, we have
\begin{equation}
{{S}_{q}}(\hat{\rho }_{i}^{tot})\le {{S}_{q}}(\hat{\rho }_{f}^{tot})
\end{equation}
On the other hand
\begin{equation}
\begin{aligned}
& {{S}_{q}}(\hat{\rho }_{f}^{tot})=-tr(\hat{\rho }_{k}^{qM{{B}_{M}}})H({{p}_{k}})-\sum\limits_{k}{tr(\hat{\rho }_{k}^{qM{{B}_{M}}})tr(\hat{E}_{k}^{\frac{q}{2}}\hat{\rho }_{i}^{qS{{B}_{S}}}\hat{E}_{k}^{\frac{q}{2}}{{\ln }_{q}}\sqrt{{{{\hat{E}}}_{k}}}\hat{\rho }_{i}^{qS{{B}_{S}}}\sqrt{{{{\hat{E}}}_{k}}})} \\ 
& +tr(\hat{\rho }_{k}^{qM{{B}_{M}}})H({{p}_{k}})+\sum\limits_{k}{p_{k}^{q}{{S}_{q}}(\hat{\rho }_{k}^{M{{B}_{M}}})}+(1-q)\sum\limits_{k}{{{S}_{q}}({{{\hat{M}}}_{k}}\hat{\rho }_{i}^{S{{B}_{S}}}\hat{M}_{k}^{\dagger }){{S}_{q}}(\hat{\rho }_{k}^{M{{B}_{M}}})} \\ 
\end{aligned}
\end{equation}
and
\begin{equation}
\begin{aligned}
& {{S}_{q}}(\hat{\rho }_{i}^{tot})={{S}_{q}}(\hat{\rho }_{q,0}^{M})tr(\hat{\rho }_{q}^{q{{B}_{M}}})tr({{{\hat{\rho }}}^{qS{{B}_{S}}}}) \\ 
& +{{S}_{q}}(\hat{\rho }_{q}^{{{B}_{M}}})tr(\hat{\rho }_{q,0}^{qM})tr({{{\hat{\rho }}}^{qS{{B}_{S}}}})+{{S}_{q}}({{{\hat{\rho }}}^{qS{{B}_{S}}}})tr(\hat{\rho }_{q}^{qM{{B}_{M}}}) \\ 
& +(1-q)\left[ {{S}_{q}}(\hat{\rho }_{q,0}^{M}){{S}_{q}}(\hat{\rho }_{q}^{{{B}_{M}}})tr({{{\hat{\rho }}}^{qS{{B}_{S}}}})+{{S}_{q}}(\hat{\rho }_{q}^{M{{B}_{M}}}){{S}_{q}}({{{\hat{\rho }}}^{S{{B}_{S}}}}) \right] \\ 
\end{aligned}
\end{equation}
From the definition of the QC-mutual information content and equation (58)(59), we obtain
\begin{equation}
\begin{aligned}
& \sum\limits_{k}{p_{k}^{q}{{S}_{q}}(\hat{\rho }_{k}^{M{{B}_{M}}})}-{{S}_{q}}(\hat{\rho }_{q,0}^{M})tr(\hat{\rho }_{q}^{q{{B}_{M}}})tr({{{\hat{\rho }}}^{qS{{B}_{S}}}}) \\ 
& +{{S}_{q}}(\hat{\rho }_{q}^{{{B}_{M}}})tr(\hat{\rho }_{q,0}^{qM})tr({{{\hat{\rho }}}^{qS{{B}_{S}}}})+(1-q)\left[ {{S}_{q}}(\hat{\rho }_{q,0}^{M}){{S}_{q}}(\hat{\rho }_{q}^{{{B}_{M}}})tr({{{\hat{\rho }}}^{qS{{B}_{S}}}}) \right] \\ 
& +(1-q)\left[ {{S}_{q}}(\hat{\rho }_{q}^{M{{B}_{M}}}){{S}_{q}}({{{\hat{\rho }}}^{S{{B}_{S}}}}) \right]+(1-q)\sum\limits_{k}{{{S}_{q}}({{{\hat{M}}}_{k}}\hat{\rho }_{i}^{S{{B}_{S}}}\hat{M}_{k}^{\dagger }){{S}_{q}}(\hat{\rho }_{k}^{M{{B}_{M}}})} \\ 
& \ge {I_{q}^{QC} }-tr(\hat{\rho }_{k}^{qM{{B}_{M}}})H({{p}_{k}}) \\ 
\end{aligned}
\end{equation}
By using Klein’s inequality, we have
\begin{equation}
\begin{aligned}
& -\sum\limits_{k}{p_{k}^{q}tr(\hat{\rho }_{k}^{M{{B}_{M}}}{{\ln }_{q}}\hat{\rho }_{k}^{M}\otimes \hat{\rho }_{q}^{{{B}_{M}}})}-{{S}_{q}}(\hat{\rho }_{q,0}^{M})tr(\hat{\rho }_{q}^{q{{B}_{M}}})tr({{{\hat{\rho }}}^{qS{{B}_{S}}}}) \\ 
& +{{S}_{q}}(\hat{\rho }_{q}^{{{B}_{M}}})tr(\hat{\rho }_{q,0}^{qM})tr({{{\hat{\rho }}}^{qS{{B}_{S}}}})+(1-q)\left[ {{S}_{q}}(\hat{\rho }_{q,0}^{M}){{S}_{q}}(\hat{\rho }_{q}^{{{B}_{M}}})tr({{{\hat{\rho }}}^{qS{{B}_{S}}}}) \right] \\ 
& +(1-q)\left[ {{S}_{q}}(\hat{\rho }_{q}^{M{{B}_{M}}}){{S}_{q}}({{{\hat{\rho }}}^{S{{B}_{S}}}}) \right]+(1-q)\sum\limits_{k}{{{S}_{q}}({{{\hat{M}}}_{k}}\hat{\rho }_{i}^{S{{B}_{S}}}\hat{M}_{k}^{\dagger }){{S}_{q}}(\hat{\rho }_{k}^{M{{B}_{M}}})} \\ 
& \ge {I_{q}^{QC} }-tr(\hat{\rho }_{k}^{qM{{B}_{M}}})H({{p}_{k}}) \\ 
\end{aligned}
\end{equation}
We define the change in the averaged free energy due to the measurement as
\begin{equation}
\Delta F_{meas}^{M}\equiv \sum\limits_{k}{p_{k}^{q}F_{k}^{M}tr(\hat{\rho }_{q,k}^{qM{{B}_{M}}})}-F_{0}^{M}tr(\hat{\rho }_{q,0}^{qM})
\end{equation}
On the other hand, the ensemble average of
work performed on M during the measurement as
\begin{equation}
\begin{aligned}
& W_{meas}^{M}\equiv \sum\limits_{k}{p_{k}^{q}\left[ \frac{{{\beta }'}}{Z_{k}^{M(1-q)}}tr(\rho _{k}^{qM{{B}_{M}}}H_{k}^{M})+\frac{{{\beta }'}}{{{Z}^{B(1-q)}}}tr({{\rho }^{qM{{B}_{M}}}}{{H}^{{{B}_{M}}}}) \right]} \\ 
& -\left[ \frac{{{\beta }'}}{Z_{0}^{M(1-q)}}tr(\rho _{q,0}^{qM}H_{0}^{M})+\frac{{{\beta }'}}{{{Z}^{B(1-q)}}}tr(\rho _{q}^{q{{B}_{M}}}{{H}^{{{B}_{M}}}}) \right] \\ 
\end{aligned}
\end{equation}
we finally obtain
\begin{equation}
\begin{aligned}
& W_{meas}^{M}\ge \frac{\beta }{\bar{Z}_{q}^{M(1-q)}}\Delta F_{meas}^{M}-\frac{\beta }{\bar{Z}_{q}^{M\left( 1-q \right)}}\Delta U_{q}^{M} \\ 
& +\sum\limits_{k}{(1-q)tr(\hat{\rho }_{k}^{qM{{B}_{M}}})({{\ln }_{q}}\hat{\rho }_{q,k}^{M})({{\ln }_{q}}\hat{\rho }_{q}^{{{B}_{M}}})} \\ 
& +(1-q)\left[ {{S}_{q}}(\hat{\rho }_{q,0}^{M}){{S}_{q}}(\hat{\rho }_{q}^{{{B}_{M}}})tr({{{\hat{\rho }}}^{qS{{B}_{S}}}}) \right] \\ 
& +(1-q)\left[ {{S}_{q}}(\hat{\rho }_{q}^{M{{B}_{M}}}){{S}_{q}}({{{\hat{\rho }}}^{S{{B}_{S}}}}) \right] \\ 
& +(1-q)\sum\limits_{k}{{{S}_{q}}({{{\hat{M}}}_{k}}\hat{\rho }_{i}^{S{{B}_{S}}}\hat{M}_{k}^{\dagger }){{S}_{q}}(\hat{\rho }_{k}^{M{{B}_{M}}})} \\ 
& +{I_{q}^{QC} }-tr(\hat{\rho }_{k}^{qM{{B}_{M}}})H({{p}_{k}}) \\ 
\end{aligned}
\end{equation}
In particular, if the free-energy difference
$\Delta {{F}^{S}}$ of the controlled system S is zero By summing up inequalities (36)(51) and (64), we obtain
\begin{equation}
\begin{aligned}
& W_{ext}^{SM}\equiv W_{ext}^{S}-W_{meas}^{M}-W_{eras}^{M}\le  \\ 
& {-\sum\limits_{m}{\frac{{{\beta }_{m}}}{{{{{\beta }'}}_{m}}}\left[ tr(\rho _{i}^{q})-tr(\rho _{f}^{q}) \right]\left( F_{mq}^{B}-U_{mq}^{B} \right)}}/{tr(\rho _{q}^{q{{B}_{1}}})...tr(\rho _{q}^{q{{B}_{n}}})}\; \\ 
& +{(1-q)\left[ {{S}_{q}}({{\rho }_{{{B}_{1}}}})...{{S}_{q}}({{\rho }_{{{B}_{m}}}}) \right]\left( {{S}_{q}}(\rho _{f}^{s})-{{S}_{q}}(\rho _{i}^{s}) \right)}/{tr(\rho _{q}^{q{{B}_{1}}})...tr(\rho _{q}^{q{{B}_{n}}})}\; \\ 
& -\sum\limits_{k}{p_{k}^{q}(1-q)({{\ln }_{q}}{{p}_{k}}){{S}_{q}}(\hat{\rho }_{q,can,k}^{M})} \\ 
& -\frac{1}{tr(\hat{\rho }_{q}^{qB})}\left\{ \frac{\beta }{{{\beta }'}}\left[ tr(\rho _{i}^{qM})-tr(\rho _{f}^{qM}) \right]\left( F_{q}^{B}-U_{q}^{B} \right) \right\} \\ 
& -\frac{1}{tr(\hat{\rho }_{q}^{qB})}\left\{ +(1-q)\left( {{S}_{q}}(\rho _{q}^{B}) \right)\left( {{S}_{q}}(\rho _{f}^{M})-{{S}_{q}}(\rho _{i}^{M}) \right) \right\} \\ 
& -\sum\limits_{k}{(1-q)tr(\hat{\rho }_{k}^{qM{{B}_{M}}})({{\ln }_{q}}\hat{\rho }_{q,k}^{M})({{\ln }_{q}}\hat{\rho }_{q}^{{{B}_{M}}})} \\ 
& -(1-q)\left[ {{S}_{q}}(\hat{\rho }_{q,0}^{M}){{S}_{q}}(\hat{\rho }_{q}^{{{B}_{M}}})tr({{{\hat{\rho }}}^{qS{{B}_{S}}}}) \right] \\ 
& -(1-q)\left[ {{S}_{q}}(\hat{\rho }_{q}^{M{{B}_{M}}}){{S}_{q}}({{{\hat{\rho }}}^{S{{B}_{S}}}}) \right] \\ 
& -(1-q)\sum\limits_{k}{{{S}_{q}}({{{\hat{M}}}_{k}}\hat{\rho }_{i}^{S{{B}_{S}}}\hat{M}_{k}^{\dagger }){{S}_{q}}(\hat{\rho }_{k}^{M{{B}_{M}}})} \\ 
\end{aligned}
\end{equation}
which implies that the work that can be extracted from the total system of S and M cannot be positive. Therefore, the conventional second law of thermodynamics is satisfied for the total system. 

\section{CONCLUSION}
The QC-mutual information introduced in section 5 characterizes the upper bound of the additional work that can be extracted from heat engines with the assistance of feedback control, or Maxwell’s demon. By using the general results in section 6 and 7, we can essentially reconcile
Maxwell’s demon with the second law of thermodynamics, which leads to a novel and general physical picture of the resolution of the paradox of Maxwell’s demon.
Moreover, our results in section 6 and 7 can be regarded as the generalizations of the second law of thermodynamics to information processing processes: feedback, measurement, and information erasure.


\begin{thebibliography}{}
	
	\bibitem{Ref1}
	Takahiro Sagawa, second law of thermodynamics with discrete quantum feedback control, Physical review Letter, 100, 080403(2008).
	
	\bibitem{Ref2}
	Rastegin, Alexey E, Some general properties of unified entropies, Journal of Statistical Physics, 143, 1120(2011). 
	
	\bibitem{Ref3}
	Abe Sumiyoshi, Okamoto Yuko, Nonextensive statistical mechanics and its applications, Springer Science  Business Media, (2001). 
	
	\bibitem{Ref4}
	L. Szilard, Z. Phys. 53, 840 (1929).
	
	\bibitem{Ref5}
	R. Landauer, IBM J. Res. Develop. 5, 183 (1961).
	
	\bibitem{Ref6}
	B. Piechocinska, Phys. Rev. A 61, 062314 (2000).
	
	\bibitem{Ref7}
	M. A. Nielsen, C. M. Caves, B. Schumacher, and H. Barnum, Proc. R. Soc. London A, 454, 277 (1998).
	
	\bibitem{Ref8}
	T. Sagawa and M. Ueda, e-Print: cond-mat/0609085
	(2006).
	
	
	
	
	
	
	
	
	
	
\end{thebibliography}
\end{document}